\documentstyle[11pt,newpasp,twoside,epsf]{article} 
\def\edcomment#1{\iffalse\marginpar{\raggedright\sl#1\/}\else\relax\fi} 
\reversemarginpar 

\begin{document} 
\title{Primordial or Dynamical Mass Segregation in Young LMC Clusters?}

\author{Richard de Grijs, Gerry Gilmore, and}
\affil{Institute of Astronomy, University of Cambridge, Madingley Road,
Cambridge CB3 0HA, UK}
\author{Rachel Johnson}
\affil{European Southern Observatory, Casilla 19001, Santiago 19, Chile} 

\begin{abstract} 
We present the detailed analysis of {\sl Hubble Space Telescope}
observations of the spatial distributions of different stellar species
in two young compact star clusters in the Large Magellanic Cloud, NGC
1805 and NGC 1818.  Based on a comparison of the characteristic
relaxation times in their cores and at their half-mass radii with the
observed degree of mass segregation, it is most likely that significant
primordial mass segregation was present in both clusters, particularly
in NGC 1805. Both clusters were likely formed with very similar initial
mass functions.
\end{abstract}

\section{Strong mass segregation on short time-scales}

One of the major uncertainties in modern astrophysics is the issue of
whether the stellar initial mass function (IMF) is universal or,
alternatively, determined by environmental effects.  Galactic globular
clusters and rich, compact Magellanic Cloud star clusters are ideal
laboratories for providing strong constraints on the universality of the
IMF, in particular because they are essentially single age, single
metallicity systems for which statistically significant samples of
individual stars over a range of masses can easily be resolved. 

Although the standard picture, in which stars in dense clusters evolve
rapidly towards a state of energy equipartition through stellar
encounters -- with the corresponding mass segregation -- is generally
accepted, observations of various degrees of mass segregation in very
young star clusters (e.g., de Grijs et al.  2002a,b and references
therein) suggest that at least some of this effect is related to the
process of star and star cluster formation itself.  Quantifying the
degree of actual mass segregation is thus crucial for the interpretation
of observational luminosity and mass functions (LFs/MFs) in terms of the
IMF, even for very young star clusters.

Therefore, we obtained F555W and F814W {\sl Hubble Space
Telescope/WFPC2} imaging observations of two young compact LMC clusters,
NGC 1805 ($\sim 10$ Myr) and NGC 1818 ($\sim 25$ Myr), covering a large
range of radii (see de Grijs et al.  2002a for observational details). 

The radial dependence of the LF and MF slopes indicate clear mass
segregation in both clusters at radii $r \la 3-6 R_{\rm core}$ (de Grijs
et al.  2002a,b).  Within the uncertainties, we cannot claim that the
slopes of the outer MFs in NGC 1805 and NGC 1818 are significantly
different, which therefore implies that these clusters must have had
very similar IMFs. 

In Fig.  1a we show the dependence of the cluster core radius on the
adopted magnitude (mass) range.  For both clusters we clearly detect the
effects of mass segregation for stars with masses $\log (m/M_\odot) \ga
0.2$ ($m \ga 1.6 M_\odot$).  It is also clear that stars with masses
$\log (m/M_\odot) \ga 0.4 \, (m \ga 2.5 M_\odot)$ show a similar
concentration, while a trend of increasing core radius with decreasing
mass (increasing magnitude) is apparent for lower masses. 

We estimate that the NGC 1818 cluster core is between $\sim 5-30$
crossing times old, so that dynamical mass segregation in its core
should be well under way.  By applying scaling laws to NGC 1805, we
conclude that its core is likely $\la 3-4$ crossing times old.  However,
since strong mass segregation is observed out to $\sim 6 R_{\rm core}$
and $\sim 3 R_{\rm core}$ in NGC 1805 and NGC 1818, respectively, for
stellar masses in excess of $\sim 2.5 M_\odot$, it is most likely that
significant primordial mass segregation was present in both clusters,
particularly in NGC 1805 (cf. Fig. 1b).

\begin{figure}
\plottwo{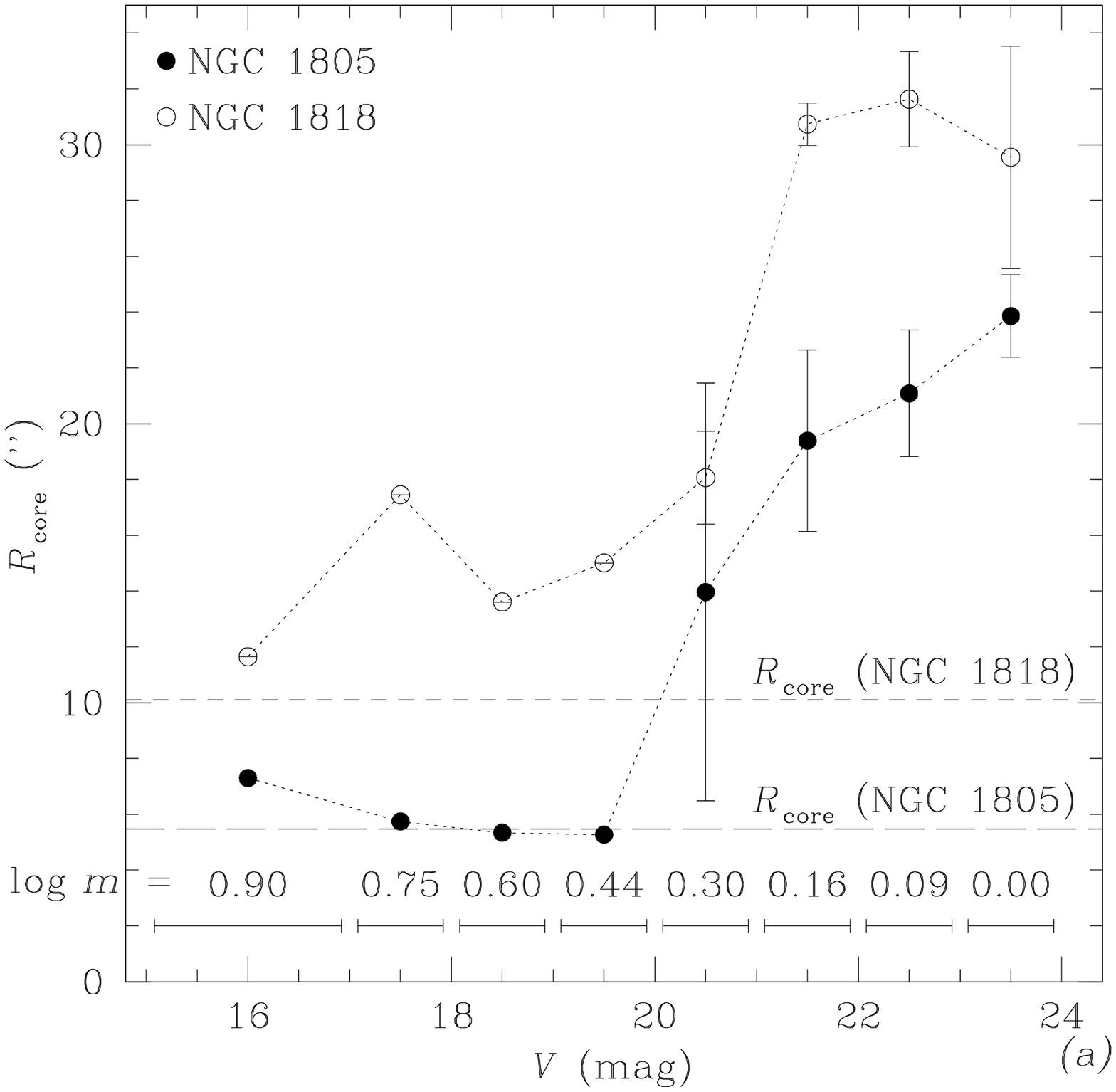}{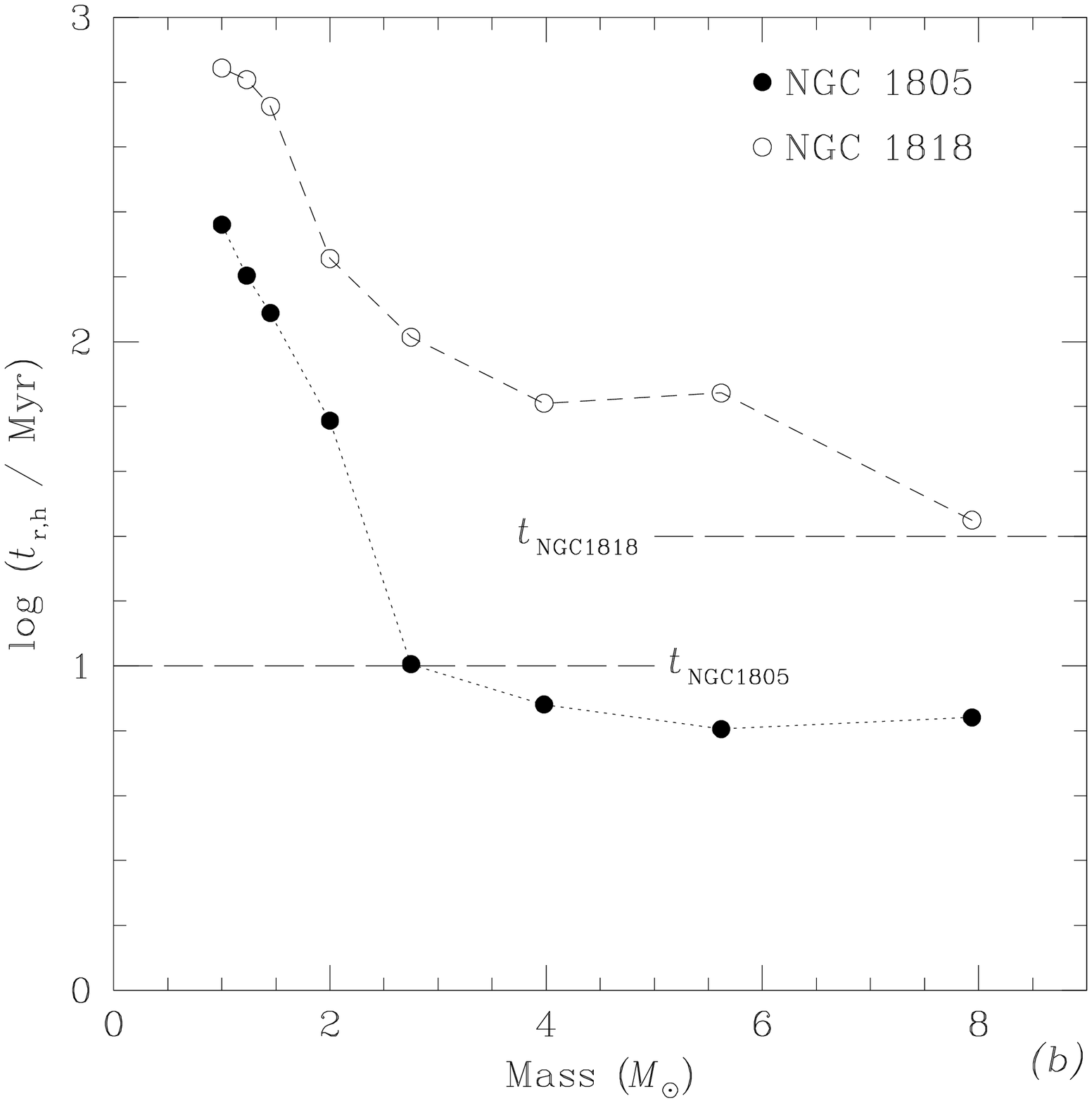}
\caption{{\it (a)} -- Core radii as a function of magnitude (mass) for
both clusters.  The error bars are driven by uncertainties in the
background subtraction; fitting ranges are indicated at the bottom of
the panel.  The horizontal dashed lines represent the overall core radii
from the clusters' surface brightness profiles.  {\it (b)} -- Half-mass
relaxation time as a function of mass for NGC 1805 and NGC 1818.  The
best age estimates for both clusters are indicated by horizontal dashed
lines.}
\end{figure}

\end{document}